\pdfoutput=1
\documentclass[letterpaper,conference]{IEEEtran}

\newif\ifpagenumbers
\pagenumberstrue

\usepackage[T1]{fontenc}
\usepackage[nocompress]{cite}
\usepackage{graphicx}
\usepackage{booktabs}
\usepackage{float}
\usepackage{array}
\usepackage{amsmath}
\usepackage{xcolor}
\usepackage{url}
\usepackage{xr-hyper}
\usepackage[hidelinks]{hyperref}
\usepackage{pdfpages}
\usepackage{refcount}

\newcommand{\papertitle}{Same Seed, Different Top Prediction: Cross-Platform Reproducibility of AlphaFold~3 Inference on TPUs, GPUs, and CPUs}

\newcommand{\AF}{AlphaFold~3}

\newsavebox{\fitbox}
\newlength{\fitwidth}
\newcommand{\fit}[3]{%
  \sbox{\fitbox}{#3}%
  \ifdim\wd\fitbox>#1\relax\sbox{\fitbox}{\resizebox{#1}{!}{\usebox{\fitbox}}}\fi
  \ifdim\dimexpr\ht\fitbox+\dp\fitbox\relax>#2\relax\sbox{\fitbox}{\resizebox*{!}{#2}{\usebox{\fitbox}}}\fi
  \global\fitwidth=\wd\fitbox
  \usebox{\fitbox}}
\newcommand{\fitcol}[1]{\fit{\columnwidth}{0.9\textheight}{#1}}
\newcommand{\fitpage}[1]{\fit{\textwidth}{0.9\textheight}{#1}}
\newcommand{\tabpairleft}{0.508\textwidth}
\newcommand{\tabpairright}{0.478\textwidth}
\newcommand{\tabpairsep}{\setlength{\tabcolsep}{3.5pt}}
\newcommand{\fitline}[1]{\fit{\linewidth}{0.9\textheight}{#1}}

\newcommand{\tablenotestyle}{\footnotesize\leftskip=0pt\rightskip=0pt\parfillskip=0pt plus 1fil\parindent=0pt\relax}
\newif\iftablenotefull
\newcommand{\tablenote}[1]{\par\vspace{2pt}%
  \iftablenotefull\setlength{\fitwidth}{\linewidth}\fi
  \ifdim\fitwidth>0pt\else\setlength{\fitwidth}{\linewidth}\fi
  {\centering\parbox{\fitwidth}{\tablenotestyle #1}\par}\global\fitwidth=0pt}
\newcommand{\isctablenote}[1]{\IfFileExists{assets/isc/tables/#1_note.tex}{\tablenote{\input{assets/isc/tables/#1_note.tex}}}{}}
\newcommand{\iscfig}[1]{assets/isc/figures/#1.pdf}

\newcommand{\citeosf}{\cite{osfregistration}}
\newcommand{\citeosfaddenda}{\cite{osfaddendumlarge,osfaddendum,osfaddendumwhy,osfaddendumf32,osfaddendumrcx,osfaddendumhlo,osfaddendumdtp,osfaddendumtri,osfaddendumrcf,osfaddendumaut,osfaddendumxvm}}

\newcommand{\capFigForest}{Top-1 flips (a) and success flips (b) between pairs of platforms, with 95\% bootstrap intervals over targets. Bands give recompilation alone on the platform with the larger rate. For FU4 and FU12, the band is recompilation on the L4 in the main study, which used seeds 1 to 5 and the default precision. FU is the follow-up. In FU1, success is a DockQ score of interface quality~\cite{basu2016dockq,mirabello2024dockq2} of at least 0.23.}

\newcommand{\capFigPerturbationApp}{Exploratory. Top-1 flip rate of nine contrasts against the median shift in ranking score, divided by the median top-two margin. The dashed line is a two-sample model. All points use seeds 1 to 5, and points 4, 6, and 7 take seeds 3 to 5 from Follow-up 12. Bars are 95\% bootstrap intervals over targets.}
\newcommand{\capFigSeeds}{Follow-up 8, outside the pre-registered study. Difference in the success rate of the overall top-ranked sample over the first n seeds, v5e minus L4, in points, with 95\% bootstrap intervals over 60 targets. The band is the $\pm$2 point criterion of Q3.}

\newcommand{\capTabArms}{Arms of the Main Study and of Follow-Ups 3 and 9}
\newcommand{\capTabHeadline}{Flips and Structure Differences in the Main Study}
\newcommand{\capTabQtwo}{Time, Cost, and Memory per Structure (Q2)}

\newcommand{\capTabSFuSixteen}{Follow-up 16: the compiled programs of the v5e and of the v6e before the Pairformer, for inputs padded to 256 tokens}

\newcommand{\capFigFuThirteen}{(a) Top-1 flip rate on the v6e against the size of a random perturbation of the trunk output, in units of $u = 2^{-8}$ (seeds 1 and 2). Triangles and squares on the x axis mark the differences measured between platforms at the trunk output, in the single and in the pair representation (layer-capture arm, 10 pilot targets). The band spans the 95\% intervals of the two primary pairs. The dashed line is recompilation on the L4 with seeds 1 to 5. (b) Top-1 flip rates with the bfloat16 trunk and with the float32 trunk (seeds 1 to 5), on a square-root scale. All rates use 60 targets and 95\% intervals over targets.}

\newcommand{\capTabBeyond}{The Difference Beyond the Main Study}
\newcommand{\capTabRepro}{The Cost of Reproducibility per Platform}

\hypersetup{pdftitle={\papertitle},pdfauthor={Lorenzo Pazienza}}\makeatletter\def\hyper@linkfile#1#2#3{#1}\makeatother

\newcommand{\Sref}[1]{\hyperlink{supplement.\getpagerefnumber{#1}}{\ref*{#1}}}
\begin{document}
\bstctlcite{BSTcontrol}

\title{\papertitle}

\author{\IEEEauthorblockN{Lorenzo Pazienza}
\IEEEauthorblockA{LUISS Guido Carli University\\Rome, Italy\\lorenzo.pazienza@gmail.com}}

\maketitle
\ifpagenumbers
\makeatletter
\def\ps@paperpagenumber{\def\@oddhead{}\def\@evenhead{}\def\@oddfoot{\hfil\footnotesize\thepage\hfil}\let\@evenfoot\@oddfoot}
\makeatother
\pagestyle{paperpagenumber}\thispagestyle{paperpagenumber}
\fi

\begin{abstract}
AlphaFold~3 returns the highest-ranked of several sampled structures and computes each prediction on one
device. Computing centers may schedule predictions on different GPU types or CPU nodes. Floating-point
results can differ between devices, so the same input and seed may return another structure. Neither AlphaFold~3's documentation nor its outputs warn that the returned structure can depend on the device.
We measure how often the device changes the returned structure, and tell users what to report and how
to make devices agree.
We used identical inputs, weights, seeds, and settings on two Google tensor processing unit (TPU) generations (v5e,
v6e), on three NVIDIA GPUs (L4, A100, H100), and on an Intel Xeon CPU.
On 60 protein-ligand complexes, with five seeds each, switching TPU generations changed the top-ranked sample
in 18.0\% of 300 predictions. Recompiling never changed the top-ranked sample on either TPU, but did so in 7.3\% on the L4
GPU.
Yet the structures barely moved. The top-ranked pose switched between correct and incorrect in 3.3\% to 3.7\% of
predictions, against about 10\% for a new seed.
These switches nearly always came from one validity check, with the ligand within 2~\AA{}.
Follow-ups found the effect again on 72 new targets and between the TPUs and the CPU. AlphaFold~3's developers verified the model's accuracy on the A100 and on the H100. Between these two GPUs, the top-ranked sample changed beyond recompilation, but the success of the top-ranked sample did not differ beyond recompilation.
An exploratory analysis links the changes to near-tied candidates.
We recommend reporting the device and every near-tied sample. A float32 trunk with highest-precision matrix products made all five
accelerators agree on nearly every prediction.
\end{abstract}

\begin{IEEEkeywords}
AlphaFold~3, protein structure prediction, high performance computing, reproducibility of results, floating-point arithmetic
\end{IEEEkeywords}

\section{Introduction}\label{sec:intro}

\AF{} predicts the joint structure of proteins, nucleic acids, and ligands~\cite{abramson2024af3}.
For each input and random seed, \AF{} samples five candidate structures and ranks them by a confidence-based score. The main output is the candidate with the highest score across all seeds~\cite{af3output}. The public implementation
computes each prediction on a single device~\cite{af3performance}, and its developers recommend scaling by running many
inputs in parallel rather than by splitting one prediction~\cite{af3issue230}. High-performance computing (HPC) centers
run many single-device predictions as independent jobs, and the scheduler may place them on different GPU types or on
CPU nodes. Of the 28 HPC-center guides for running \AF{} that we read, 23 ask for one device per job, and seven let the same
job script run on any of several GPU types, or suggest CPU nodes when GPUs are busy~\cite{hpctacc,hpcsherlock}
(Supplementary Section~\Sref{S-sec:s-guides}).

Floating-point results depend on the order of operations~\cite{goldberg1991floating}. In parallel reductions, that order depends on the architecture~\cite{ahrens2015reproblas,collange2015reduction}, so results can depend on the platform, which we define as the compiler and the hardware together. This dependence on the platform is an established concern in HPC~\cite{miao2023expression,wen2025mindthegap}. For users, what matters is whether \AF{} returns a different answer, not whether the coordinates differ in the last bits. \AF{} can rank a different candidate first, and that candidate may place the ligand correctly on one platform but not on another. \AF{}'s developers verified its numerical accuracy on the NVIDIA A100 and H100
GPUs~\cite{af3installation}, but the documentation does not say whether the two GPUs return the same top-ranked sample. Neither \AF{}'s logs nor its output files warn that the results depend on the device. A user can receive a
different structure from the same input and seed without knowing it. We measure how often the platform changes the structure returned by \AF{}, and we compare this rate with two
references, recompilation and a new seed. We also tell users what to report and how to make platforms agree. No work we reviewed measures such changes across TPUs, GPUs, and CPUs, or compares them with recompilation and with a new seed (Section~\ref{sec:background}).

The data freeze records 20\,983 \AF{} processes on 230 targets, one fresh process per run, with identical inputs, weights, seeds, and settings. The six platforms were two Google tensor processing unit (TPU) generations (v5e and v6e), three NVIDIA GPUs (L4, A100, and H100), and an Intel Xeon (Ice Lake) CPU host. On each accelerator, we repeated the default
runs in a separate process that compiles the model again, so that the effect of recompilation is measured, not assumed.
The core comparison covered the TPUs, the L4, and the CPU. We pre-registered the core comparison on the Open Science Framework (OSF)~\citeosf{} after a first pilot and before the first main-study run. The A100 and H100 were added later as follow-ups. All follow-ups had a written plan before they ran, and eleven were also registered as OSF addenda~\citeosfaddenda{}.

Changing only the TPU generation led to a different top-ranked sample of a seed (a top-1 flip) in about one in five predictions, where a prediction is one target with one seed. Recompiling on either TPU never caused a flip. The
TPU v6e and the L4 GPU differed about as often, and recompiling on the L4 GPU alone caused fewer top-1 flips. A replication on 72 new targets confirmed the effect. In follow-ups, the platform also exceeded recompilation between the H100 and A100 GPUs, between the TPUs and the CPU, and with both the TPU v6e and the L4 GPU at their highest matrix-product precision. Yet the predicted structures barely moved. We count a pose as correct when its ligand lies within
2~\AA{} (1~\AA{} = 0.1~nm) of the experimental position and passes validity checks. The top-ranked pose rarely switched
between correct and incorrect, and a new seed did so about three times as often. The platform adds a smaller source of variation than the seed, but one that nobody chose or reported. A benchmark at five seeds can still differ by several points. At five seeds, the
success rate of \AF{}'s main output differed by up to 6.7 points between platforms (the largest of six differences, with an interval not adjusted for that selection) and
by 5.0 points under recompilation on the L4 alone (Section~\ref{sec:discussion}). On the 72 targets of Follow-up~11, the same changes of platform moved this success rate by at most 1.4 points (Supplementary Section~\Sref{S-sec:s-campaign-sim}).

An exploratory analysis suggests why: the two best candidates often score almost equally, so rounding differences can swap them. Follow-ups tie the differences to the network's bfloat16 arithmetic together with the default matrix-product precision. With a float32 trunk at the highest matrix-product precision, the five accelerators agreed on
nearly every prediction and had the same per-seed success rate. Neither setting alone sufficed between the v6e and the L4 (Follow-ups~12 and~17). The exact operation remains open. We measured single-device runs on cloud VMs, not a computing-center campaign.

This paper makes four contributions.
\begin{itemize}
\item A measurement design that separates three causes of a changed result (the platform, recompilation,
and the seed), under rules fixed in advance (Section~\ref{sec:method}).
\item The effect on the returned prediction, its boundaries, its exploratory relation to the ranking margin, and a float32-trunk mitigation at the highest matrix-product precision
(Section~\ref{sec:results}).
\item Decision data for HPC users on the cost of reproducibility, on ranking rules, on campaigns that mix devices, and on what to
report and pin (Sections~\ref{sec:results-cost} and~\ref{sec:discussion}).
\item A released record of inputs, outputs, and a data freeze, with a 7-line port of DeepMind's code to Cloud TPUs
and with a run harness (Data and Code Availability).
\end{itemize}

\section{Background and Related Work}\label{sec:background}

\subsection{What a Seed Fixes in \AF{}}
\AF{} predicts a structure in two stages. A trunk built mainly from 48 Pairformer blocks turns the input and its sequence alignments into two representations. The single representation has one vector per token, and the pair representation has one vector per pair of tokens. There is one token per standard amino acid or nucleotide, and one per heavy atom for ligands and modified residues~\cite{abramson2024af3}. A diffusion module then samples atom coordinates from these
representations, and a confidence head evaluates the result. By default \AF{} draws five samples, or candidate structures,
per seed and orders them by a ranking score built from predicted confidences and penalty terms. \AF{}'s main output is the top-ranked sample across seeds~\cite{af3output}. A seed fixes the random numbers of the sampling, but not the arithmetic. On another device, the same random numbers pass through differently compiled code, so the values can
differ.

\subsection{Why Platforms Differ}\label{sec:why-differ}
Floating-point addition is not associative: the same numbers summed in another order can give a different result~\cite{goldberg1991floating}. JAX compiles the model with XLA, which does not fix the order of the terms of a sum. XLA has one backend for each kind of chip, and each backend may choose its own order~\cite{xla_opsem,stablehlo_spec}. Precision defaults also differ by backend. \AF{} v3.0.4, the version we ran, computes the trunk in bfloat16 and the diffusion module in float32 on every platform~\cite{af3repo}, but when it sets no precision for a float32 matrix product, the backend applies a default precision~\cite{jaxprecision}. That default is single-pass bfloat16 on the TPUs, TensorFloat-32 (TF32) on the GPUs, and IEEE float32 on the CPU, with 8 bits, 11 bits, and 24 bits of significand. The highest matrix-product precision runs float32 matrix products as six-pass bfloat16 on the TPUs and as IEEE float32 on the
GPUs. On all three GPUs, \AF{} computes the gated linear units with a Triton kernel from Tokamax~\cite{tokamax}. Triton is a language for writing GPU kernels~\cite{tillet2019triton}. Results can also change when the same program is compiled
again in a new process. We call this second compilation recompilation. On GPUs, XLA can select kernels by timing candidates at compile
time, unless a flag requests deterministic operations~\cite{xla_proto}. Upstream \AF{} later documented same-seed
differences between separately compiled processes~\cite{af3performancev305}. On an A100, a user traced a structure
change under another XLA reduction policy to 12 early bfloat16 values. A maintainer replied that \AF{} is
numerically sensitive and that such changes are expected~\cite{af3issue772}. Issue \#772 studies three selected
outliers and gives no frequency. In issue \#694, a user reported differences of up to 0.25~\AA{} in the predicted aligned error between two runs with the same seed, and a maintainer answered that such outputs should be identical~\cite{af3issue694}.

We use reproducibility in the computational sense, which means obtaining consistent results from the same input data, code, computational
steps, and conditions of analysis~\cite{nasem2019reproducibility}. We keep \AF{}'s inputs, weights, seeds, and settings
and vary one condition on purpose. We either move to another platform and so change the hardware together with its compiler
and default settings, or we only compile the model again on the same platform. We assess reproducibility at three levels of tolerance, which require bitwise-identical outputs, the same top-ranked sample, or the same success. We test these levels under three changes of increasing size, which are rerunning the compiled program, recompiling on the same platform, and changing the platform. In ACM's terms, our same-platform reruns test repeatability. A change of platform by the same team has no ACM term~\cite{acm2020badging}.

\subsection{Related Work}
In HPC, reproducible summation makes parallel reductions independent of their
order~\cite{ahrens2015reproblas,collange2015reduction}, and vendor libraries offer conditional reproducibility
modes~\cite{intel2025cnr}. Ensemble consistency tests decide whether a climate model's output on a new platform is
statistically distinguishable from an ensemble run on the original platform~\cite{baker2015new}. Miao et al.\ vary compiler settings on a single machine and isolate the
expressions in GPU code that cause the resulting inconsistencies~\cite{miao2023expression}. We study the consequence of
numerical error for a scientific decision, in a compiled machine-learning application, across accelerator families.
In deep learning, training variance from seeds and tooling is well
documented~\cite{pham2020variance,summers2021nondeterminism}, and Zhuang et al.\ compared this variance on several GPUs and on a
TPU~\cite{zhuang2022randomness}. Atomic reductions make results vary between runs on one GPU~\cite{shanmugavelu2024impacts}. Neural-network inference differs across 75 CPU and GPU platforms and, through
autotuning, between sessions on one GPU~\cite{schloegl2023deviations}. RepDL makes training and inference bitwise
reproducible on CPUs and GPUs~\cite{xie2025repdl}. Greedy decoding of large language models diverges across GPU types,
GPU counts, and batch sizes when rounding changes the order of near-tied
tokens~\cite{yuan2025understanding,cooper2026accelerating,he2025defeating}, a mechanism close to the one we describe. These works
concern training, general inference, or text generation, and none of them measures the ranked output of a scientific model across
TPUs, GPUs, and CPUs or compares this output with recompilation and with a change of seed.

For structure prediction, PoseBusters provides the protein-ligand benchmark and the validity checks we
use~\cite{buttenschoen2024posebusters}. Runs N' Poses studies how co-folding models generalize to new
complexes~\cite{skrinjar2026cofolding}. Systems work has made AlphaFold2-family training and inference faster and
larger~\cite{cheng2024fastfold,zhu2024scalefold}, and MegaFold brought the training of \AF{}-style models to ISC High
Performance~\cite{la2026megafold}. An earlier preprint mapped AlphaFold2 queries across the chips of a TPU v5e
slice~\cite{pazienza2026accelerator}. AlphaFold2 confidence has been compared on CPU and GPU for six proteins~\cite{chaudhary2026openomics}. On one GPU type, the confidence scores of Protenix-v2 varied between three runs with fixed seeds on five nanobody-antigen complexes, and deterministic PyTorch algorithms removed the variation but made the runs about 1.2 times as long~\cite{ramon2026beyond}. What is new here is a systematic measurement across TPUs, GPUs, and CPUs under fixed weights, inputs, and seeds. We compare each change of platform with recompilation and with a change of seed. We also measure the effect on the returned pose. Finally, we test a float32 intervention.

\section{Method}\label{sec:method}

\subsection{Platforms, Arms, and Runs}\label{sec:port}
We ran DeepMind's public \AF{}~\cite{abramson2024af3} v3.0.4 (commit 85c4d20) with its official model weights~\cite{af3installation}, a single
file that we identify by its SHA-256 hash (Supplementary Section~\Sref{S-sec:s-arms}). Upstream \AF{} officially supports NVIDIA GPUs and documents unofficial CPU-only and Apple Silicon modes~\cite{af3installation}. We added the TPU as a backend with a 7-line patch that removes no lines. The patch requires the portable \texttt{xla} attention on the TPU and changes nothing else in \AF{}.
The TPUs use jax and jaxlib 0.10.2 with libtpu 0.0.43.2, and the GPUs and the CPU use \AF{}'s pinned software stack, with jax 0.10.2 and, on the GPUs, the CUDA~12 plugins.

\begin{table}[t]
\caption{\capTabArms}\label{tab:arms}
\centering\small\fitcol{\begin{tabular*}{\linewidth}{@{\extracolsep{\fill}}llr@{}}
\toprule
arm & differs from base & USD/h \\
\midrule
v5e & base & 1.56 \\
v6e & base & 2.97 \\
L4 & base & 0.85/0.90 \\
L4 Triton & Triton attention kernel & 0.85/0.90 \\
L4 deterministic & deterministic-operations flag & 0.85/0.90 \\
v6e highest & highest matrix-product precision & 2.97 \\
L4 highest & highest matrix-product precision & 0.85 \\
CPU & base & 1.15 \\
A100 & base & 5.03/5.54 \\
A100 Triton & Triton attention kernel & 5.03 \\
A100 highest & highest matrix-product precision & 5.03 \\
H100 & base & 10.98 \\
\bottomrule
\end{tabular*}%
}Each variant arm differs from its device's base arm in one setting. USD/h is the on-demand list price, and two prices mean regions with different prices. Regions without a price read use the europe-west4 price (Supplementary Section~\Sref{S-sec:s-main-q2}). The A100 and the H100 have 80 GB of memory.

\end{table}
\label{sec:arms}An arm is one device with its settings. Table~\ref{tab:arms} lists the arms of the main study and of Follow-ups~3 and~9. Section~\ref{sec:prereg} describes the arms of the later follow-ups, and Supplementary Table~\Sref{S-tab:s-runs} gives the
runs of each study through Follow-up~12. The TPUs are the TPU v5e (v5e) and the TPU v6e (v6e). The GPUs are the NVIDIA L4, A100, and H100, and the CPU is an Intel Xeon (Ice Lake) host. Base arms use \AF{}'s portable \texttt{xla} attention. The Triton arms (L4 and A100) use Triton attention, which is \AF{}'s default on GPUs. The highest arms (v6e, L4, and A100) run float32 matrix products at JAX's highest matrix-product precision, and the L4 deterministic arm sets XLA's deterministic-operations flag.
We added the L4 deterministic arm after a pilot diagnostic and before the main study. Each combination of platform, target, seed, and
repetition runs as its own fresh process with an empty compilation cache. Every inference run used a Google Cloud
virtual machine (VM) with one device. Supplementary Section~\Sref{S-sec:s-arms} lists the machine type, host CPU,
zones, and VM placement of every arm. Repetitions r1 and r2 of a target run in separate processes, so each repetition compiles the model from scratch. The two repetitions ran on one VM, except in Follow-ups~6, 9, 15, 18, and~19 and in the TPU recompilations of Follow-ups~13, 14, and~17 (Supplementary Section~\Sref{S-sec:s-arms}). The only exception to the empty cache is a warm rerun, which loads from a cache the model that r1 compiled. In the main study, r1 has a warm rerun for seed~1 on the 10 pilot targets (a subset of the 60).

\subsection{Targets, Inputs, and Metrics}
The main study used 60 protein-ligand complexes from PoseBusters V2,
a benchmark of 308 complexes~\cite{buttenschoen2024posebusters,posebusterszenodo}.
A fixed rule kept the 132 complexes with one protein chain, one ligand in \AF{}'s component dictionary, and at
most 1024 tokens.
\AF{} pads each input to a token bucket, one of the padded input sizes it compiles for~\cite{af3performance}.
We drew the 60 targets with a fixed seed, as evenly as possible across the buckets.
There are 19 targets of up to 256 tokens, 19 of up to 512 tokens, 14 of up to 768 tokens, and 8 of up
to 1024 tokens.
In the base arms except the CPU, each complex ran with five seeds.
\AF{} was trained on PDB structures released up to 30~September~2021~\cite{abramson2024af3}, so its training data
may include 21 of our 60 targets. Such overlap can raise success rates. On the 39 later targets, the top-1 flip rate was
18.5\% between the TPUs and 22.6\% between the v6e and the L4 (Supplementary Section~\Sref{S-sec:s-cutoff}).
The selection did not use release dates.
Alignments came from the ColabFold server~\cite{mirdita2022colabfold} and were frozen. Inputs carry no templates, and
each ligand is given by its Chemical Component Dictionary code, without cofactors.

For each sample we read the coordinates, the ranking score, and two confidences. The confidences are the mean predicted local distance difference test (pLDDT) and the interface predicted template modeling score (ipTM). A ligand pose is the
placement of the ligand in the binding pocket. We count a sample as successful when its ligand pose lies within 2~\AA{} of
the experimental pose and passes the PoseBusters checks for docked poses~\cite{buttenschoen2024posebusters}
(posebusters 0.6.5~\cite{posebusterscode}). The distance is the symmetry-corrected heavy-atom root-mean-square deviation (RMSD), computed with RDKit~\cite{rdkit}, after superposition on the C-alpha atoms of the pocket, defined as the protein residues with any heavy atom within 10~\AA{} of the experimental ligand.
A sample that cannot be scored counts as no success. A prediction is one target with one seed on one arm. Each prediction returns five samples. The top-ranked sample of a prediction is the one of its five samples with the highest ranking score, read at full precision. The prediction succeeds if this sample does. \AF{} applies the same rule across all seeds to choose its main
output~\cite{af3output}. The pre-registered statistics count per-seed choices, because each seed has its own top-ranked sample. The ranking margin is the score of the top-ranked sample minus the score of the next. We compare each sample with the
sample of the same target, seed, and index under the other condition. We record whether the coordinates of the two samples are
identical as \AF{} writes them (three decimals in the mmCIF file), and call such samples bit-identical. Warm reruns and the runs with the flag on the A100 and on the H100 were checked file by file for identical files.
For each pair we measure two shifts. The ligand shift is the RMSD of the ligand atoms after the two binding pockets are superposed. The protein shift is the RMSD of the C-alpha atoms after the two proteins are superposed. We also measure the absolute change in the ranking score and in the two confidences, pLDDT for each atom and ipTM for the interface between the molecules. Time, cost, and device memory come from the run records (Supplementary Section~\Sref{S-sec:s-main-q2}).

\subsection{Contrasts and Statistics}\label{sec:contrasts}
The plan asks three questions. Q1 asks whether the platform changes what \AF{} returns, beyond what recompilation and a change of seed do. Q2 asks for the time, cost, and device memory per structure on each platform. Q3 asks how many seeds make the success rates of the platforms agree. A contrast compares two conditions on matched predictions, which share the target and the seed. Structural measures compare the samples with the same index. Recompilation compares the two repetitions, r1 and r2, on the same platform. A platform contrast compares the first repetition, r1, on two different platforms. The seed contrast compares different seeds on one platform, as a yardstick. The top-1 flip of the seed contrast carries no information, because the sample numbers of different seeds are unrelated. For the seed contrast, only the success flip and the structural shifts are informative. On the L4, the default-attention
contrast compares the L4 Triton arm with the L4 arm. The precision contrast compares a platform with its own
highest-precision arm, and the highest-precision contrast compares two highest-precision arms. The pre-specified
primary pairs are the v5e and the v6e, the v6e and the L4, and the L4 and the CPU. Other pairs are secondary. Comparing
the v5e with the v6e changes the chip but not the compiler version or the precision policy, though the compiler generates code specific to each chip (Follow-up~16). Every other pair measures the whole platform stack
(hardware, compiler, and backend defaults), not the chip alone.

For each contrast, we compute three rates over the matched predictions. A top-1 flip occurs when two matched predictions have different top-ranked samples. A success flip occurs when one of the two predictions succeeds and the other fails. A flip that matters is a top-1 flip that is also a success flip. In such a flip, the change of condition alters both the returned sample and whether that sample succeeds.
Each rate is the share of matched predictions with that flip. We also report medians and 90th percentiles of the
structural and confidence differences, and the share of matched samples that are bit-identical. We report 95\% percentile intervals from a paired bootstrap over targets, with 10\,000 replicates. The bootstrap uses its own random seed, 2026. An interval with no event is [0.0, 0.0]. An exact binomial bound over targets remains valid when the seeds of a target are correlated. The bound gives a one-sided 95\% upper limit of 4.9\% when none of 60 targets has an event, and of 25.9\% when none of 10 has one. The key quantity is the rate between two platforms minus the rate of recompilation on the first-named platform, and the bootstrap computes both rates on the same resampled targets. The plan fixed this reference before any run. ``Beyond recompilation'' means that the interval of this difference lies above~0. An interval that includes 0 means that the difference cannot be told apart from recompilation. The reading of an interval that includes 0 is not a test of equivalence. With a zero reference and events of one sign,
the interval lies above 0 once the events fall on at least 4 of 60 targets. Because recompilation on the TPUs never
changed the top-ranked sample, we also report every pair with a GPU against the larger recompilation rate of its two platforms, which is the more conservative reference. We say which reference each number uses. A rule fixed before any analysis compares the highest-precision contrast with recompilation. If the interval of the highest-precision contrast minus recompilation includes 0, the plan attributes the platform difference to the default matrix-product precision. Otherwise, the plan attributes the difference to the platform as a whole, which includes the chip, the compiler, and the L4's gated-linear-unit kernel. We do not separate these parts. The answer to Q3 is the smallest number of seeds $n$, from 1 to 5, at which the platforms agree within $\pm 2$ percentage points. The platforms agree when, for every pair of platforms, the difference in the success rate of the main output has an interval within $\pm 2$ points. Following benchmarking
guidelines~\cite{hoefler2015scientific}, we state whether each platform's measurements are deterministic before comparing platforms. We apply no correction for multiple comparisons, and we read secondary and
post hoc results as descriptive.

\subsection{Pre-Registration, Follow-Ups, and Campaign}\label{sec:prereg}
We registered the plan on OSF~\citeosf{} after a first pilot and before the first main-study run, and logged every
later change with its date and reason (Supplementary Section~\Sref{S-sec:s-corrections}). We ran a pilot on 10 targets to check the pipeline and to size the study. We do not report the pilot's outcomes as findings. After reading the main results, we
specified twenty-one follow-up experiments, which test whether the result holds under other conditions. Follow-ups~3
and~9 change the GPU generation, and Follow-up~4 uses a CPU. Follow-ups 6, 10, and 11 use inputs from \AF{}'s own data
pipeline, larger inputs, and new targets. Follow-ups 7 and 12 use an older software stack and the highest matrix-product precision.
Follow-ups 1, 2, 5, and 8 check protein-protein complexes, the trunk outputs, the host's input preparation, and more
seeds. Follow-up~13 tests why the top-ranked sample changes, and Follow-up~14 tests whether a float32 trunk makes every accelerator agree. Follow-up~16 compares the compiled programs of the two TPUs. Follow-up~17 separates the trunk's
data type from the precision, and Follow-up~18 uses Triton attention on the H100 and on the A100. Follow-ups~15 and~19 measure recompilation of the float32 trunk, and Follow-up~15 also measures recompilation of the L4 Triton arm. Follow-up~20 measures recompilation of the L4 without autotuning, and Follow-up~21 measures recompilation on new VMs. For each follow-up, we wrote the question and the design before it ran. A reading rule, written before a run, fixes how its result is interpreted. Follow-up~3, the highest-precision contrast of Follow-up~4, and every follow-up from Follow-up~5 on, except Follow-up~9, had such a rule (Supplementary
Table~\Sref{S-tab:followups-isc}, and below for Follow-up~13). The follow-ups are outside the pre-registered study and
reported whatever they showed. Follow-ups 10, 11, and 13 to 21 also have OSF addenda registered before they ran~\citeosfaddenda{}. Supplementary Section~\Sref{S-sec:s-corrections} lists where the follow-ups departed from their addenda.

\label{sec:fu13-design}Follow-up~13 asks why the top-ranked sample flips and what preventing the flips would cost. Our run harness adds each
change, and \AF{}'s code stays unchanged. The perturbation arm adds standard normal noise, scaled to a relative size
$\epsilon$, to the two outputs of the trunk on the v6e (60 targets, seeds 1 and 2). We give $\epsilon$ in units of $u = 2^{-8}$, the largest relative rounding error of bfloat16, and use $\epsilon = 0$, $u/64$, $u/8$, $u$, $2u$, $4u$, and $8u$. The layer-capture arm records the single representation after each of the 48 Pairformer blocks and the pair representation after every second block on a token subset. The layer-capture arm runs on the v5e, on the v6e, and twice on the L4 (10 pilot targets, seed 1). The
float32-trunk arm runs the trunk in float32 with the highest matrix-product precision on the v6e and twice on the L4
(60 targets, seeds 1 to 5). The float32-trunk arm changes \AF{}'s \texttt{bfloat16} setting. The setting also covers the input embedding and the confidence head's Pairformer. The precision setting covers every float32 matrix product without an explicit precision.
For this arm, $D_F$ is the difference between two platforms beyond recompilation with the float32 trunk, minus the same difference with the bfloat16 trunk at the default precision. The arm reads ``reduced'' if the interval of $D_F$ lies below 0 (Supplementary Section~\Sref{S-sec:s-fu13}). Follow-up~14 runs the float32 trunk also on the v5e, on the A100,
and on the H100, and the deterministic flag on the A100 and on the H100 (Supplementary Section~\Sref{S-sec:s-fu14}).
Follow-up~15 recompiles the float32 trunk on the A100 and on the H100 (10 pilot targets, seeds 1 to 5) and the L4 Triton arm on 60 targets (Supplementary Section~\Sref{S-sec:s-fu15}).

\label{sec:campaign}The data freeze records 20\,983 \AF{} processes on 230 targets (60 of the main study, 30 of Follow-up~1, 68 of Follow-up~10, and 72 of Follow-up~11), with 104\,895 samples. Of the 20\,983 processes in the freeze, only 4 failed, and all 4 were planned out-of-memory stops of the memory-limit measurement. The first layer-capture session of Follow-up~13, whose 20 processes failed at the first capture, is not in the freeze, and we relaunched the session after a fix.
Supplementary Table~\Sref{S-tab:s-provisioning} gives the provisioning of every
arm through Follow-up~12. Some runs
stopped because of a preemption or a setup failure, and others could not start for lack of capacity. We relaunched
all of these runs with the same model settings, some on other VMs (Supplementary Sections~\Sref{S-sec:s-relaunches}
and~\Sref{S-sec:s-provisioning-changes}). On 27~October~2026 the Python patch release changed (3.12.15 to 3.12.16). We could check the effect of this change only on the v6e, where the outputs stayed identical (Supplementary Section~\Sref{S-sec:s-arms}).
We dropped no run and chose no run based on its results. We made a data freeze on 25~November~2026. The freeze pins down the locally held files of every run in this paper with one identifier (Data and Code Availability). Three post hoc analyses came later and read the frozen files. The reading of Follow-up~3 that section 11 of the plan fixes dates from 26~November~2026. The analyses of the main output with the float32 trunk and of the tie rule on the targets of Follow-up~11 date from 27~November~2026.

\section{Results}\label{sec:results}

\subsection{The Top-Ranked Sample Depends on the Platform}\label{sec:results-decision}
\begin{table*}[t]
\caption{\capTabHeadline}\label{tab:headline}
\centering\footnotesize\fitpage{\begin{tabular}{@{}l@{\hspace{10.4pt}}r@{\hspace{10.4pt}}l@{\hspace{10.4pt}}l@{\hspace{10.4pt}}l@{\hspace{10.4pt}}l@{\hspace{10.4pt}}l@{\hspace{10.4pt}}l@{\hspace{10.4pt}}l@{}}
\toprule
contrast & pairs & top-1 (\%) & success (\%) & matter (\%) & ligand (\AA{}) & C-alpha (\AA{}) & identical & returned \\
\midrule
\multicolumn{9}{@{}l}{\emph{Recompilation alone}} \\
Recompile, v5e & 300 & 0.0 [0.0, 0.0] & 0.0 [0.0, 0.0] & 0.0 [0.0, 0.0] & 0 & 0 & 1500 of 1500 & none \\
Recompile, v6e & 300 & 0.0 [0.0, 0.0] & 0.0 [0.0, 0.0] & 0.0 [0.0, 0.0] & 0 & 0 & 1500 of 1500 & none \\
Recompile, L4 & 300 & 7.3 [4.7, 10.3] & 1.7 [0.3, 3.3] & 1.0 [0.0, 2.3] & 0.00390 & 0.00760 & 35 of 1500 & 0.18 \\
\multicolumn{9}{@{}l}{\emph{Primary pairs}} \\
v5e vs v6e & 300 & 18.0 [14.3, 21.7] & 3.7 [1.7, 5.7] & 3.0 [1.3, 5.0] & 0.0165 & 0.0445 & 0 of 1500 & 0.18 \\
v6e vs L4 & 300 & 20.7 [16.3, 25.3] & 3.3 [1.0, 6.0] & 2.7 [0.7, 5.0] & 0.0288 & 0.0766 & 0 of 1500 & 0.17 \\
\quad minus recompile L4 (points) & 300 & +13.3 [8.7, 18.0] & +1.7 [$-$0.7, 4.3] & +1.7 [$-$0.3, 4.0] & n/a & n/a & n/a & n/a \\
L4 vs CPU & 12 & 8.3 [0.0, 25.0] & 16.7 [0.0, 33.3] & 8.3 [0.0, 25.0] & 0.0231 & 0.0882 & 0 of 60 & 0.32 \\
\multicolumn{9}{@{}l}{\emph{Change of seed (reference)}} \\
Seeds, v6e & 600 & 80.5 [77.5, 83.2] & 9.7 [5.0, 15.0] & 7.7 [3.8, 12.0] & 0.186 & 1.05 & 0 of 3000 & n/a \\
\multicolumn{9}{@{}l}{\emph{Default GPU attention}} \\
Default attention, L4 & 300 & 13.7 [9.7, 17.7] & 2.7 [1.0, 4.3] & 1.7 [0.3, 3.3] & 0.00660 & 0.0280 & 0 of 1500 & 0.16 \\
\bottomrule
\end{tabular}
}Flips are percentages of predictions, with 95\% bootstrap intervals over targets. The indented row subtracts L4 recompilation, the secondary reference. The registered reference is recompilation on the v6e, which had no events, so against it the excess equals the rate. Ligand and C-alpha are median RMSDs of paired samples. Returned (post hoc) is the median ligand RMSD between differing returned samples. The top-1 flip of the seed row carries no information. None means no change, and n/a means not applicable. The L4~vs~CPU row covers the 6 targets that the CPU ran, with seeds 1 and 2. On these targets, the top-1 rate of this row minus L4 recompilation, with seeds 1 to 5, is +5.0 [0.0, 15.0] points. Follow-up 4 covers all 60 targets.

\end{table*}

\label{sec:results-checks}All 2992 \AF{} processes of the main study exited normally, and every process that compiled the model
started from an empty compilation cache. Every warm rerun was identical to its first run in
every output file except the timestamp of each structure file. At full precision, no top-ranked sample tied with the second-ranked sample, so no flip comes from breaking an exact tie. The success criterion cannot be applied to three targets with a sugar (glycan) ligand, because \AF{} removes one atom from these ligands by default. As the plan specifies, these targets count as no success on every platform. Supplementary Table~\Sref{S-tab:s-glycans} gives two sensitivity variants, one of them added after we found the problem.

Each TPU and the L4 ran the 60 targets with five seeds, so each of these platforms made 300 predictions. The base arms of the TPUs and of the L4 use \AF{}'s portable \texttt{xla} attention, and the TPUs run through our 7-line patch (Section~\ref{sec:port}). Although the two TPUs ran the same compiler stack, inputs, weights, and seeds, the top-ranked sample changed between the v5e and the v6e in 54 of the 300 predictions, 18.0\% with a 95\% interval of [14.3, 21.7]. On each TPU, we also ran every prediction twice, in two processes that compiled the model separately. The top-ranked sample never changed on either TPU, and all 1500 samples were bit-identical (Table~\ref{tab:headline}).
Follow-up~5 built the model inputs on the four host CPU families of all arms except the H100. The inputs were identical in all 26\,550 comparisons (the 60 targets and the 30 of Follow-up~1, five seeds), so the differences do not come from the host's input preparation on these four host CPU families. The host CPU of the H100 arm was not tested. The second primary pair is the v6e and the L4. Between these two platforms, the top-ranked sample changed in 20.7\% [16.3, 25.3] of predictions. On the L4, unlike on the TPUs, recompilation alone changed the top-ranked sample in 7.3\% [4.7, 10.3] of predictions. The difference between the two platforms exceeds this rate by 13.3 percentage points [8.7, 18.0]. \AF{}'s main output is the top-ranked of the 25 samples of the five seeds. In a post hoc count, the main output changed on 16 of the 60 targets between the TPUs and on 25 between the v6e and the L4. The success of the main output changed on 3 and 4 of the 60 targets (5.0\% and 6.7\%), and on 3 under recompilation on the L4. When the top-ranked sample changed, the two returned poses lay a median 0.18~\AA{} apart
between the TPUs and 0.17~\AA{} between the v6e and the L4, as under recompilation on the L4 (0.18~\AA{}). The returned poses lay more than 2~\AA{} apart in 1 of the 53 and in 3 of the 60 changed predictions outside the glycan targets (Supplementary
Table~\Sref{S-tab:s-winners}, post hoc). The third primary pair, the L4 and the CPU, ran on 6 targets in the main study
and on 60 in Follow-up~4 (Section~\ref{sec:results-general}).

In a success flip, the top-ranked sample succeeds on one platform and fails on the other. Success flips occurred in 3.7\% [1.7, 5.7] of predictions between the TPUs and in 3.3\% [1.0, 6.0] between the v6e and the L4. In these two pairs, top-1 flips were five to six times more frequent than success flips, and flips that matter were rarer still (Table~\ref{tab:headline}). Over all pairs in Fig.~\ref{fig:forest}, top-1 flips ranged from 12.0\% to 30.0\%, and success flips ranged from 0.7\% to 5.3\%. Between the v6e and the L4, the registered rule subtracts recompilation on the v6e, which is zero. By that rule, success flips and flips that matter were beyond recompilation by 3.3 points [1.0, 6.0] and 2.7 points [0.7, 5.0]. Neither rate could be told apart from recompilation on the L4 (Table~\ref{tab:headline}).

\begin{figure*}[t]
\centering
\includegraphics{\iscfig{fig2_forest}}
\caption{\capFigForest}
\label{fig:forest}
\end{figure*}

\subsection{Structures and Confidences Move Little}\label{sec:results-structure}
No sample was bit-identical in any pair of platforms. Between the TPUs, the ligand of paired samples moved by a median
RMSD of 0.0165~\AA{}, and a change of seed on the v6e moved it by 0.186~\AA{}, about 11 times more
(Table~\ref{tab:headline}; Supplementary Fig.~\Sref{S-fig:s-ecdf}). A few samples moved more. At least 2~\AA{} separated 13 of 1425 sample pairs between the TPUs and 20 between the v6e and the L4, and the largest difference was 8.8~\AA{} (post hoc). The median ranking-score difference was 0.000321 between the TPUs and 0.000648 between the v6e and the L4. The shift between the TPUs is about a quarter of the median ranking margin (0.0012), and the shift between the v6e and the L4 is about half of that margin. Both shifts are enough to swap the two best samples when the margin is small.

\subsection{Why the Top-Ranked Sample Changes}\label{sec:results-mechanism}
We specified an exploratory margin analysis after reading the main results. In this analysis, the top-1 flip rate fell as the ranking margin grew. We split the predictions into five equal groups by margin. Between the v5e and the v6e, the top-ranked sample changed in 38.3\% [28.3, 48.9] of the predictions in the group with the smallest margins and in 6.7\% [1.5, 13.7] of the predictions in the group with the largest margins (Supplementary Fig.~\Sref{S-fig:s-margin-isc}). Across nine contrasts, the flip rate
rose with the median ranking-score difference between the two conditions divided by the median top-two margin. Platform pairs and arithmetic changes on a single platform follow the same rising trend. Recompilation on the L4
shifted the scores by 0.08 times the margin and flipped 7.3\% of predictions. The v6e and the L4 Triton differed by 0.55
times the margin and flipped 22.3\% (Fig.~\ref{fig:app-perturbation}). A simple model, with nothing fitted, approximates these rates from the two medians. In the model, the gap between the two best scores follows an exponential distribution, and a change of condition adds independent normal noise to each score. The top-ranked sample changes when the noise reverses the gap. The model predicts fewer flips than we measured between the two TPUs and for three changes of arithmetic on the L4, which are recompilation, the deterministic flag, and Triton attention (points 1 to 3 of Fig.~\ref{fig:app-perturbation}, Supplementary
Section~\Sref{S-sec:s-exploratory}). The relation is descriptive, not causal, and it does not explain every flip. Flips also occur at large margins, in 13.3\% [4.8, 21.2] of predictions in the largest-margin quintile between the v6e and the L4. Over the nine contrasts, with the three contrasts that involve a highest-precision arm at seeds 1 and 2, 33 of 39 flips that matter kept both top-ranked ligands within 2~\AA{} of the experimental pose. In these 33 flips, only one PoseBusters check changed, the minimum distance to the protein. The same held for 39 of the 41 success flips in the six contrasts of Supplementary Table~\Sref{S-tab:s-pbflips}, from the main study and from Follow-ups~3 and~9 (post hoc).
In the other six flips that matter, the ligand crossed the 2~\AA{} threshold, and only PDB entry 7N4W changed its binding mode
(7.1~\AA{} against 1.2~\AA{}). Between the H100 and the A100, the returned poses lay a median 0.16~\AA{} apart, and 1 of 59 lay more than 2~\AA{} apart (Supplementary Section~\Sref{S-sec:s-winners}). In a post hoc analysis, we computed the ligand RMSD of each returned structure from the experimental pose. In every contrast, the two RMSDs differed by a median of at most 0.1~\AA{}. Between the v6e and the L4, the mean difference over the changed predictions was $-0.13$~\AA{} [$-0.44$, 0.03], almost all of it from two predictions of 7N4W. The two-decimal ipTM of the output files had a median difference of 0 (Supplementary
Section~\Sref{S-sec:s-returned}).

\label{sec:results-fu13}Follow-ups~13 and~14 (Section~\ref{sec:fu13-design}) passed every pre-specified check
(Supplementary Sections~\Sref{S-sec:s-fu13} and~\Sref{S-sec:s-fu14}).

\begin{figure}[t]
\centering
\includegraphics{\iscfig{fig_fu13}}
\caption{\capFigFuThirteen}
\label{fig:fu13}
\end{figure}

With the trunk in float32, the v6e and the L4 returned a different top-ranked sample in 0.3\% [0.0, 1.0] of
predictions, and recompilation on the L4 never changed the top-ranked sample (Fig.~\ref{fig:fu13}(b)). Beyond recompilation on the L4, the platform difference fell from 13.3 points to 0.3 points. The change of the platform difference, $D_F$, was $-$13.0 points [$-$17.7, $-$8.3], and the pre-specified rule reads it as ``reduced''. Follow-up~14 ran the float32 trunk also on the v5e, on the A100, and on the H100. All 10 pairs of the five accelerators then agreed by the pre-specified rule, which asks for an upper 95\% bound below 5\% and was set after Follow-up~13 was read. The top-1 flip rates of these pairs were 0.3\% to 2.0\%, with no upper bound above 3.7\% (four pairs in Fig.~\ref{fig:fu13}(b), all in Supplementary Section~\Sref{S-sec:s-fu14}). Against the bfloat16 trunk, the float32 trunk changed the
returned sample in 15.7\% to 25.0\% of predictions on the five platforms. The success rate of the float32 trunk was 70.7\% on all five platforms, within the 70.0\% to 71.0\% of the bfloat16 trunk. Follow-up~17 ran a float32 trunk at each backend's default precision, which on the TPU runs float32 products as single-pass bfloat16. The v6e and the L4 then differed in 20.0\% [16.3, 23.7] of predictions, and pairs of GPUs differed in 6.7\% to 8.3\%. No pair of the four accelerators agreed by the rule (Supplementary Table~\Sref{S-tab:s-fu17-pairs}). Between the v6e and the L4, only the two changes together
reduced the difference, an interaction between the settings.

From $u/64$ to 1~$u$, a random perturbation of the trunk output on the v6e changed the returned sample in 5.0\% to 10.0\% of predictions (Fig.~\ref{fig:fu13}(a)). Even the smallest size, $u/64$, swapped some near-tied samples. Recompilation on the L4 changed 7.3\% of predictions, measured on seeds 1 to 5 instead of 1 and 2. From 2~$u$, the rate reached the level between platforms. On seeds 1 and 2, both primary pairs correspond to a perturbation of about 2~$u$. The 95\% intervals of the two pairs run from $u/64$, the smallest size and a censored bound, to 4.1~$u$. At the trunk output, the differences measured between the platforms are about 1~$u$ in the single representation and 4~$u$ to 9~$u$ in the pair representation. The real differences are thus comparable to the perturbation of about 2~$u$, or larger. For top-1 flips, the perturbation is consistent with the near-tie
account.

In the layer-capture arm, the platforms already differed at the input of the Pairformer on every pilot target, so the
differences arise earlier in the trunk (Supplementary Section~\Sref{S-sec:s-fu13}). Likewise, in issue \#772, 12 early bfloat16 values decided one structure on an A100~\cite{af3issue772}. Follow-up~16 showed that the v5e and the v6e compile the same program into different ones, already before the Pairformer (Table~\ref{tab:app-hlo}). This registered, descriptive comparison does not show which difference changes the numbers.

\subsection{What the Difference Survives}\label{sec:results-robust}
\begin{table*}[t]
\centering
\begin{minipage}[t]{\tabpairleft}\centering
\caption{\capTabBeyond}\label{tab:beyond}
\small\tabpairsep\fitline{\begin{tabular}{@{}rlll@{}}
\toprule
FU & Pair & Top-1 flip & Beyond \\
 &  & (\%) & recompilation \\
 &  &  & (points) \\
\midrule
1 & v6e vs L4, protein-protein & 30.0 [22.0, 38.0] & +10.7 [2.7, 18.7] \\
3 & L4 vs A100 & 12.0 [8.3, 16.0] & +2.0 [$-$1.7, 5.7] \\
4 & v5e vs CPU & 17.5 [10.8, 24.2] & = rate \\
4 & v6e vs CPU & 15.8 [9.2, 23.3] & = rate \\
4 & L4 vs CPU & 19.2 [12.5, 26.7] & +11.8 [5.0, 19.0] \\
6 & v6e vs L4, official inputs & 26.7 [21.7, 31.3] & +18.7 [13.3, 24.0] \\
9 & H100 vs A100 & 20.7 [16.3, 25.0] & +10.7 [6.3, 15.0] \\
11 & v5e vs v6e, new targets & 18.3 [14.7, 22.2] & = rate \\
11 & v6e vs L4, new targets & 21.9 [18.1, 25.8] & +16.7 [12.8, 20.8] \\
15 & Recompile, L4 Triton & 7.7 [4.7, 11.0] & +0.3 [$-$3.0, 4.0] \\
\bottomrule
\end{tabular}
}The table gives the top-1 flip and its excess over the larger recompilation rate of the pair, with 95\% intervals over targets. FU is the follow-up. ``= rate'' means a recompilation rate of 0. A CPU pair uses the other platform's rate, because the CPU ran once. FU15 compares recompilation of the L4 Triton arm with recompilation of the L4 arm. FU4 subtracts L4 recompilation on seeds 1 to 5. On seeds 1 and 2, the excess over L4 recompilation is +14.2 [7.5, 20.8] (post hoc).

\end{minipage}\hfill
\begin{minipage}[t]{\tabpairright}\centering
\caption{\capTabRepro}\label{tab:repro}
\small\tabpairsep\fitline{\begin{tabular}{@{}lrrrrrr@{}}
\toprule
 & default & \multicolumn{2}{c}{deterministic flag} & \multicolumn{3}{c}{float32 trunk} \\
\cmidrule(lr){2-2}\cmidrule(lr){3-4}\cmidrule(l){5-7}
platform & flip & flip & time & flip & time & memory \\
\midrule
\multicolumn{7}{@{}l}{\emph{Recompilation on one platform}} \\
v5e & 0.0 & n/a & n/a & 0.0* & 1.51 & 0.99 \\
v6e & 0.0 & n/a & n/a & 0.0* & 1.58 & 1.02 \\
L4 & 7.3 & 0.0 & 1.03--1.12 & 0.0 & 1.55 & 1.24 \\
A100 & 10.0 (8.0*) & 0.0* & 1.15* & 0.3 & 1.43 & 1.24 \\
H100 & 9.3 (2.0*) & 0.0* & 1.18* & 0.3 & 1.47 & 1.27 \\
\multicolumn{7}{@{}l}{\emph{Change of platform}} \\
pairs & 12.0--23.3 & 21.0 & n/a & 0.3--2.0 & n/a & n/a \\
\bottomrule
\end{tabular}
}Top-1 flips are in \% of predictions on 60 targets with seeds 1 to 5. An asterisk (*) marks values measured on 10 pilot targets, with seed 1 on the TPUs and with seeds 1 to 5 on the GPUs. Parentheses give the default on those 10 targets. Time and compiled memory are ratios of medians to the default arm. The L4 flag time is a range over buckets. In the last row, the flag column gives the v6e against the L4 deterministic arm. The last row gives ranges over the same 10 pairs of the five accelerators, with the bfloat16 trunk and with the float32 trunk. Compiled memory is the median of the bytes that the compiled program needs, as reported by XLA.

\end{minipage}
\end{table*}
With the deterministic flag, recompilation on the L4 changed no top-ranked sample (0 of 300) and left all 1500 samples bit-identical. The flag did not reduce the difference from the v6e. Between the v6e and the L4 deterministic arm, the top-ranked sample changed in 21.0\% [16.7, 25.3] of predictions, against 20.7\% without the flag. The flag changed the L4's top-ranked sample in 11.3\% [7.7, 15.7] of predictions, so it makes the L4 repeatable but changes the L4's results. Turning off XLA's
autotuning removed the recompilation differences on the L4 (300 of 300 predictions identical), consistent with kernel
selection by timing as their source (Follow-up~20).

Follow-up~12 set the highest matrix-product precision on both the v6e and the L4, with five seeds. The two platforms still differed. Top-1 flips occurred in 20.7\% [16.0, 25.7] of predictions, success flips occurred in 2.3\% [0.7, 4.3], and flips that matter occurred in 2.0\% [0.7, 3.7]. Recompilation at the highest matrix-product precision was not measured, so the reference is recompilation at the default precision. On the v6e, recompilation at the default precision gave bit-identical results. Against this reference, the pre-specified rule attributes all three rates to the platform on all 60 targets. With the two seeds of the registered analysis, success flips and flips that matter had not been distinguishable from recompilation. By that rule, these two rates were attributed to the default matrix-product precision. Adding seeds 3 to 5 in Follow-up~12 changed that reading (Supplementary Section~\Sref{S-sec:s-fu12}).
The L4 does not recompile bit-identically, so we also compared the pair with recompilation on the L4 at the default precision. Only the top-1 flip rate exceeded this reference, by 13.3 percentage points [9.3, 17.3]. For success flips and flips that matter, the excess was 0.7 points [$-$1.3, 3.0] and 1.0 points [$-$0.3, 2.7], and both intervals include zero. The setting changes only float32 matrix products and leaves the bfloat16 trunk unchanged (Section~\ref{sec:why-differ}). With five seeds, the precision setting alone changed the top-ranked sample in 20.0\% [15.0, 25.0] of predictions on the v6e and in 10.3\% [7.0, 13.7] on the L4.

\label{sec:results-general}Outside the main study (Table~\ref{tab:beyond} and Fig.~\ref{fig:forest}), the top-1 difference
exceeded recompilation in every pair of devices with the bfloat16 trunk except two. Between the L4 and the A100, the difference exceeded recompilation on the L4 by 4.7 points [0.3, 9.3] but did not exceed recompilation on the A100 (Supplementary Table~\Sref{S-tab:s-fu03-hw-minus-rc}). With Triton attention, the same held (Follow-up~18 and Supplementary Section~\Sref{S-sec:s-gtat}). Between the A100 and the CPU at the highest matrix-product precision, the reading that the plan fixes for Follow-up~3 places the difference at the recompilation level (Supplementary Table~\Sref{S-tab:s-fu03-s11}). With \AF{}'s default Triton
attention (Follow-up~18), the H100 and the A100 still exceeded recompilation, by 11.0 points [6.7, 15.3] against recompilation on the H100 and by 9.0 points [4.7, 13.7] against recompilation on the A100 (Supplementary Section~\Sref{S-sec:s-fu18}). On 72 new targets, a replication (Follow-up~11) found a top-1 difference beyond recompilation of similar size between the v5e and the v6e and between the v6e and the L4 (18.3 points and 21.9 points, against 18.0 points and 20.7 points in the main study). On these targets, four of the eight secondary readings on success flips and flips that matter were beyond
recompilation. In Follow-up~7, switching to an older software stack never changed the top-ranked sample on the v6e and changed it in 7.0\% of predictions on the L4. Both rates are below the rate of a change of platform (20.7\% between the v6e and the L4). The rate on the L4 could not be told apart from recompilation. On inputs
of 1025 to 1536 tokens, the top-1 flip exceeded recompilation on the first-named platform, which changed no top-ranked sample, by 20.6 points [14.7, 26.5] between the TPUs and by 25.3 points
[17.6, 33.5] between the v6e and the L4. Against recompilation on the L4, the v6e and the L4 exceeded recompilation by 12.4 points [7.1, 18.2]. On size controls of at most 1024 tokens, the two pairs exceeded recompilation on the first-named platform by the same 22.4 points [16.5, 28.8], and the v6e and the L4 exceeded recompilation on the L4 by 16.5 points [10.6, 22.4]. By its rule, Follow-up~10 detected no size difference at this sample size (Supplementary Table~\Sref{S-tab:followups-isc}).

\subsection{Time, Cost, Seeds, and Reporting Rules}\label{sec:results-cost}
Table~\ref{tab:app-q2} in the appendix gives time, cost, and memory per structure (Q2). A structure here is one target and seed with five samples. When the 8 targets of the 1024-token bucket share one compilation, a structure cost 0.098~USD on the v5e, 0.124~USD on the v6e, 0.159~USD on the L4, and 0.081~USD on the L4 Triton at on-demand prices. Table~\ref{tab:repro} gives
the cost of reproducibility. The deterministic flag made recompilation bit-identical on the L4 and, on 10 pilot targets,
gave identical output files on the A100 and on the H100, at a small cost in time and none in compiled memory (Supplementary Table~\Sref{S-tab:s-fu14-cost}). The float32 trunk removed nearly all differences between platforms but took about 50\% longer. With this trunk, recompiling changed the
top-ranked sample in 0 of 300 predictions on the L4, in 0 of 10 on each TPU, and in 1 of 300 on the A100 and in 1 of 300 on the H100
(Follow-ups~13 to~15 and~19). Most GPU coordinates still differed.

By the pre-specified rule (Section~\ref{sec:contrasts}), Q3 is not reached. With 60 targets, one target moves a success rate
by 1.67 points, against a margin of $\pm 2$ points (appendix, Fig.~\ref{fig:app-seeds}).

\label{sec:results-rules}After reading the results, we specified a further analysis of the frozen outputs before computing it. The reporting-rules analysis compared rules for choosing the structure to report (Supplementary Section~\Sref{S-sec:s-rules}). More
seeds did not help. With \AF{}'s own rule, the top sample across $n$ seeds, the v5e and the L4 returned different
samples for 26.4\% [23.8, 29.2] of predictions at one seed (seeds 1 to 20, Follow-up~8). For blocks of twenty seeds of a target, the two platforms differed for 48.3\% [36.7, 61.7] of blocks. A tie rule declares a tie when the top two scores differ by at most 0.001. The tie rule cut the per-seed flip between platforms to 3.0\% to 6.0\%, but it found ties in 53.7\% to 54.0\% of predictions. A unit counts as a flip under this rule only when each side reports one sample. The threshold is tuned on these protein-ligand data and should be validated on new targets. We chose 0.001 post hoc, from the median margin of 0.0012. The rule trades agreement for more structures reported.

\section{Discussion}\label{sec:discussion}

\subsection{Implications}
In many predictions, the two best samples score almost equally. Changing the platform, recompiling, setting the deterministic flag, or raising the precision shifts the scores slightly and can reverse the order of these two samples. The relation between near-tied samples and flips is exploratory (Section~\ref{sec:results-mechanism}, Fig.~\ref{fig:app-perturbation}).

This dependence on the platform matters first for HPC centers. In our cloud runs, the A100-H100 and L4-CPU pairs returned different top-ranked
samples in 20.7\% and 19.2\% of predictions, 10.7 points to 11.8 points more often than recompilation on these GPUs
(Section~\ref{sec:results-general}). A campaign that spreads its jobs over different GPU types or over CPU nodes is likely to
mix results that differ in this way. Pinning one GPU type reduces these differences but does not remove them. Without the deterministic flag, recompilation alone changed the top-ranked sample on each of the three GPUs (Table~\ref{tab:repro}). On the L4, the rate was 5.3\% to 19.3\% on the other target sets of Fig.~\ref{fig:forest}. None of the 28
HPC-center guides we read warns that the same seed can give a different result on another GPU type. Supplementary
Section~\Sref{S-sec:s-campaign-sim} estimates such campaigns by scaling the measured rates with the share of moved jobs. The simulation covers five moves of jobs, which go from the A100 to the L4, from the A100 to the H100, from the L4 to the CPU, from the v6e to the L4, and from the v5e to
the v6e. With a quarter of the jobs moved, the returned sample changed in 3.0\% to 5.5\% of one-seed jobs and in 5.4\%
to 10.4\% of whole-target jobs. A recompilation on the same GPU already changed 5.0\% (L4, seeds 1 and 2) to 10.0\% (A100) of one-seed jobs, and a recompilation on the TPUs changed none.

For structural biologists, a change of platform often changes which sample \AF{} returns for a seed, but rarely whether that sample succeeds. At five seeds, the largest of six differences in the success rate of the main output among the
v5e, v6e, L4, and L4 Triton was $-$6.7 points (v6e minus L4, [$-$13.3, $-$1.7]). Recompilation on the L4 alone moved the success rate of the main output by $-$5.0 points [$-$11.7, 0.0] (Supplementary Section~\Sref{S-sec:s-best25}). Follow-up~8 kept the point estimates of
the v5e and the L4 within 1.7 points up to 20 seeds. With the float32 trunk at the highest matrix-product precision, all five accelerators had a per-seed success rate of 70.7\%. Single-pose benchmarks should name the platform. A change of platform changes the returned sample in about one prediction in five, and the two poses then lie a median 0.16~\AA{} to 0.18~\AA{} apart.

\subsection{Recommendations}\label{sec:recs}
(1) Report the accelerator, the machine type, the software stack, and the seed. On GPUs, also report the determinism setting. The accelerator and the determinism setting changed the top-ranked sample in our runs.
For example, a methods section could include the line ``\AF{} v3.0.4 on one NVIDIA L4 (\texttt{g2-standard-8}), jax 0.10.2 with the CUDA~12 plugins, \texttt{xla} attention, deterministic flag off, seeds 1 to 5.''
(2) Do not expect more seeds to stabilize the choice. With \AF{}'s own rule, platforms disagreed more often on the
returned sample as seeds were added (Section~\ref{sec:results-rules}).
(3) For each prediction, report every sample whose score lies within 0.001 of the top score, not only the top-ranked sample. We chose this threshold post hoc on these data, and the threshold declared ties in about 54\% of predictions (Section~\ref{sec:results-rules}). On the 72 new targets of Follow-up~11, a post hoc check with the same threshold cut the per-seed flip to 1.9\% between the v5e and the v6e and to 2.5\% between the v6e and the L4, and the threshold declared ties in 55.0\% and in 59.4\% of the predictions of the two pairs (Supplementary Section~\Sref{S-sec:s-ties11}). The tie rule reduces the problem but
does not remove it. In a post hoc count, at least one of the 25 samples of a target (five seeds with five samples each) was successful for 85.0\% to 86.7\% of the targets on each of the five accelerators. In each of the four pairs we compared, the platforms disagreed on whether a successful sample existed for at most 1 of 60 targets (Supplementary Section~\Sref{S-sec:s-best25}). For the L4 and the CPU, the count covers the 10 samples of seeds 1 and 2.
(4) Rank by the full-precision score, which is \AF{}'s own rule. The two-decimal score that \AF{} also writes picked a
different top sample in 51.9\% of the main study's predictions (a post hoc count). The rounded score makes platforms agree more often only because rounding creates ties, which go to the same sample index on every platform.
(5) To get identical results from repeated runs on one GPU type with \texttt{xla} attention and an unchanged software stack (VM image, driver, CUDA libraries, and jax), run \AF{} with XLA's deterministic flag (\texttt{XLA\_FLAGS=}\allowbreak\texttt{-{}-xla\_gpu\_\allowbreak deterministic\_ops=true}; Table~\ref{tab:repro}). We tested the flag on the L4 with 60 targets, and on the A100 and on the H100 with the 10 pilot targets. In every test, both runs of a target ran on one VM. The flag made the runs 1.03 to 1.27 times as long, depending on the device and the input size, and added no compiled memory. We did not test the flag with Triton attention. The flag does not make different platforms agree. On the L4, the flag changed the returned sample in 11.3\% of predictions against the default, so set it before a campaign starts. Turning off XLA's
autotuning (\texttt{-{}-xla\_gpu\_\allowbreak autotune\_level=0}) also made recompilation on the L4 identical, and the runs took 1.01 times as long as with autotuning (Follow-up~20).
(6) Where the top-ranked sample should agree across platforms, run the trunk in float32 with the highest matrix-product precision. To do so, set \texttt{bfloat16} to \texttt{none} in \AF{}'s global model configuration. Setting \texttt{bfloat16} to \texttt{none} takes a one-line change, because no command-line flag sets the data type of the trunk. Also set the environment variable \texttt{JAX\_DEFAULT\_MATMUL\_PRECISION=highest}. Every pair of the five accelerators then returned the same top-ranked sample in 98.0\% to 99.7\% of predictions, although no sample kept identical coordinates. In a post hoc count over five seeds, the main output of the float32 trunk differed between two accelerators on at most 1 of the 60 targets, and the success of the main output never changed. The main output of the bfloat16 trunk differed between two accelerators on 15 to 27 of the 60 targets (Supplementary Section~\Sref{S-sec:s-mainout}). The float32 trunk took about 1.5 times as long (Table~\ref{tab:repro}) and up to 2.19 times as long at 1024 tokens (Supplementary Table~\Sref{S-tab:s-fu14-cost}). We measured the float32 trunk with \texttt{xla} attention on the GPUs, and we did not measure the largest input it can run. The success rate of the float32 trunk stayed within the range of the bfloat16 trunk (Section~\ref{sec:results-fu13}), so we recommend the float32 trunk only where agreement across platforms matters, not as a default.

\subsection{Limitations}\label{sec:limits}
\AF{}'s released code runs each prediction on one device. The code compiles the model for a single device, and its flag for
machines with several GPUs pins each run to one GPU. Inference across several devices would require changes to \AF{}'s
code and could add differences in reduction order. We ran on cloud VMs, not at a computing center, and we measured
devices, not a campaign.

Our attribution is partial. The host check excludes the host's input preparation. Follow-ups~13, 14, and~17 implicate
the bfloat16 arithmetic together with the default matrix-product precision, because the differences between all five
accelerators nearly vanished only with the trunk in float32 at the highest matrix-product precision
(Section~\ref{sec:results-fu13}). The float32 trunk at the highest matrix-product precision is a measured mitigation, not a located cause. Between the v6e and the L4, data type and precision reduced the differences only together (Follow-ups~12 and~17), so the exact operation
remains open. Within a platform, we did not separate the chip, the compiler, and \AF{}'s gated-linear-unit kernel for
GPUs. The GPU base arms use \texttt{xla} attention, not \AF{}'s default Triton attention. Recompiling the L4 Triton arm changed the
top-ranked sample in 7.7\% of predictions, against 7.3\% on the L4. This difference of +0.3 points [$-$3.0, 4.0] is descriptive (Follow-up~15). With Triton attention on both GPUs, the L4 and the A100 differed beyond recompilation on the L4 (Supplementary Section~\Sref{S-sec:s-gtat}, post hoc) but not beyond recompilation on the A100 (Follow-up~18, descriptive). The same held with \texttt{xla} attention (Follow-up~3). On the TPUs, \AF{} runs through our 7-line patch, which the official release lacks. The CPU arm ran once
per target, so we know CPU recompilation only from the 6 targets that ran in the repeated pilot and in the main study, on separate VMs. All 12
predictions of these 6 targets were bit-identical, which bounds the share of targets with a change at 39\% (an exact one-sided 95\% bound).

Success flips and flips that matter are rare. The two primary pairs with 60 targets have 8 to 11 of each in 300 predictions, so the intervals of these rates are wide. Several follow-up readings depend on whether the lower bound of an
interval is exactly 0.0. Post hoc bias-corrected and accelerated intervals on the same replicates would change 3 of 53
readings, all toward ``beyond''. The registered readings stand (Supplementary Section~\Sref{S-sec:s-rare}). We applied no
correction for multiple comparisons, so readings that rest on few events are best taken as hypotheses.

The main study uses 60 single-chain protein-ligand targets, with ligands given by their Chemical Component Dictionary
code, ColabFold alignments, no templates, and no cofactors. Of these targets, 21 were released on or before the training cutoff
and may have been in the training data. Besides an Intel Xeon CPU, we measured only Google TPUs and NVIDIA GPUs. Google Cloud did not offer AMD GPUs during our study, although systems such as the European LUMI use them~\cite{lumig}. \AF{}'s installation guide also asks for an NVIDIA GPU~\cite{af3installation}.

We measured \AF{} only. Open co-folding models such as Boltz and Chai~\cite{wohlwend2024boltz1,chai2024chai1} also rank sampled candidates, so near-ties can decide which candidate they return. The papers of these models do not discuss platform dependence. A Boltz-2~\cite{passaro2025boltz2}
port to Intel GPUs agrees more closely with NVIDIA in float32 than in bfloat16~\cite{boltzcommunitypr20}.
Supplementary Section~\Sref{S-sec:s-open} lists open questions.

\section{Conclusion}\label{sec:conclusion}
We asked whether the device that computes an \AF{} prediction changes its result. With the same input, weights, and seed, a change of platform altered which sample was ranked first in about one prediction in five. Recompiling never changed the top-ranked sample on the TPUs and changed it less often on the GPUs. The structures moved little. When the top-ranked sample changed, the two returned poses lay about 0.17~\AA{} apart, and their success rarely changed. The change of the top-ranked
sample, however, survived the highest matrix-product precision, the deterministic flag on the L4, new targets, larger
inputs, and a move between the A100 and the H100. In an exploratory analysis, the frequency of flips tracked the score
shift relative to the ranking margin in nine contrasts. Controlled follow-ups were consistent with this near-tie account. Only with a float32 trunk and the highest matrix-product precision together did the five accelerators return the same top-ranked sample in
nearly every prediction and have the same per-seed success rate (the per-seed success rates of the bfloat16 trunk already lay within 70.0\% to 71.0\%, and in a post hoc count the five-seed main output differed between two accelerators on at most 1 of the 60 targets). With the default settings, a benchmark at five seeds could report
success rates up to 6.7 points apart on two platforms, about as far apart as two compilations on one GPU (5.0 points). A seed is
not a complete description of an \AF{} run. Anyone who reports a seed should also report the platform and its software stack. Anyone who needs a reported result to hold should report all seeds used, the ranking margin of the chosen
sample, and every sample within 0.001 of the top score, rather than one top-ranked sample.

\section*{Data and Code Availability}\label{sec:availability}
We release the study's inputs, \AF{} outputs, and data freeze in two Zenodo records, with the code. Inputs with their
alignments, target lists, manifests, the data freeze, and run records are in record R1, DOI
10.5281/zenodo.24057470~\cite{zenodor1}, under CC~BY~4.0. \AF{} outputs, pipeline results, and derived analysis outputs
are in record R2, DOI 10.5281/zenodo.24057544~\cite{zenodor2}, under the \AF{} Output Terms of
Use~\cite{af3outputterms}, for non-commercial use. The supplementary material is appended to this paper. Both records are public. The two records include no model parameters, which are available only from Google DeepMind. The code is at \url{https://github.com/lorenzopazienza/af3-cross-platform}, in the commit tagged \texttt{zenodo-v1}, under the MIT license. A copy of the code is also in record R1. The plan, \url{https://osf.io/fegsy}, and its eleven addenda~\citeosfaddenda{} are registered
on OSF.
The data freeze of 25~November~2026 covers the locally held files of every run in this paper, and it has an identifier{},
\texttt{d706dffc07fec6d86471e28f\allowbreak 062178e4f309193d3bfd6a6\allowbreak 6de47959eb3bfdd8f}. Supplementary Section~\Sref{S-sec:s-reproduction} describes the manifest that this identifier names and the input check made before release. The post hoc analyses of 26 and 27~November~2026 read the frozen files.

\emph{Notice.} \AF{} outputs, and the tables and figures in this paper computed from them, are provided under and
subject to the \AF{} Output Terms of Use. The predicted structures, confidences, and ranking scores were not modified. The tables and figures are computed from these outputs as described in Section~\ref{sec:method}. \AF{} model parameters are not redistributed. The use of these parameters is subject to the Model Parameters Terms of Use~\cite{af3weightsterms}. The TPU runs use an
unofficial patch, and Google has neither reviewed nor endorsed this work.

\section*{Acknowledgment}
This work was supported by Google Cloud credits from the Google TPU Builders Program. The author conceived and
directed the study, made every decision, launched every run, and validated every result. The author used Claude Code
(Anthropic) to write the code and the analyses and to assist with the writing. The author reviewed the code and the
analyses at every step, worked on every sentence of the paper by hand, and is fully responsible for the content. Each
number is checked, by a script or by hand, against the versioned file it cites, and the code repository lists each number with its source.

\label{refs-start}
\bibliographystyle{IEEEtran}
\bibliography{references}

\clearpage
\appendix
\section{}
The figures and tables below are also in the supplement. These figures and tables let the reader check claims that the body states in
words.

\begin{figure}[H]
\centering
\includegraphics{\iscfig{fig4_perturbation}}
\caption{\capFigPerturbationApp}
\label{fig:app-perturbation}
\end{figure}

\begin{figure}[H]
\centering
\includegraphics{\iscfig{fig5_seed_curve}}
\caption{\capFigSeeds}
\label{fig:app-seeds}
\end{figure}

\subsection*{Time, Cost, and Memory (Q2)}
Table~\ref{tab:app-q2} gives time, cost, and memory per structure (Q2). A structure is one target and seed with five samples. Inference includes compilation, and wall time also includes loading and writing. At 1024 tokens, inference took
a median of 222~s on the v5e, 167~s on the v6e, 656~s on the L4, and 349~s on the L4 Triton. At on-demand prices, a structure cost
0.098~USD on the v5e, 0.124~USD on the v6e, 0.159~USD on the L4, and 0.081~USD on the L4 Triton. These costs are amortized, as the plan defines them: the 8 targets of the bucket share one compilation. The largest input
that ran was 1536 tokens on the v5e, 2048 tokens on the v6e, and 2560 tokens on the L4. The A100 and H100 completed at least 5120 tokens.

\subsection*{Number of Seeds (Q3)}
By the pre-specified rule (Section~\ref{sec:contrasts}), Q3 is not reached. No number of seeds from one to five met the rule.
Follow-up~8 extended the v5e and L4 runs to 20 seeds (Fig.~\ref{fig:app-seeds}). The rule then answers Q3 with 2
seeds, but the answer is not stable. The interval leaves $\pm 2$ points again at every $n$ from 3 to 20, while the
estimate stays between $-$1.7 points and +1.7 points. With 60 targets, the 95\% interval of a difference in success rate spans about
10 points (one target is 1.67 points), so it can lie within $\pm 2$ points only when it collapses to zero. The rule asked for more precision than 60 targets give. Q3 could not be answered as designed, and
we read its results descriptively.

\makeatletter\setlength{\@dblfptop}{0pt}\setlength{\@fptop}{0pt}\makeatother
\begin{table*}[!t]
\caption{\capTabQtwo}\label{tab:app-q2}
\centering\footnotesize\fitpage{\begin{tabular}{@{}lrrrrrrrrrrl@{}}
\toprule
 & \multicolumn{2}{c}{256 tokens (s)} & \multicolumn{2}{c}{1024 tokens (s)} & \multicolumn{4}{c}{USD per structure at 1024} & \multicolumn{2}{c}{GiB at 1024} & memory limit (tokens) \\
\cmidrule(lr){2-3}\cmidrule(lr){4-5}\cmidrule(lr){6-9}\cmidrule(lr){10-11}\cmidrule(l){12-12}
arm & compile & inference & compile & inference & fresh & amortized & amort.\ Spot & warm & memory & limit & largest; first failed \\
\midrule
v5e & 33 & 45 & 40 & 222 & 0.113 & 0.0981 & 0.0354 & 0.0981 & 7.28 & 15.7 & 1536; 2048 \\
v6e & 35 & 43 & 63 & 167 & 0.170 & 0.124 & 0.0746 & 0.121 & 7.28 & 31.2 & 2048; 2560 \\
L4 & 59 & 98 & 64 & 656 & 0.173 & 0.159 & 0.0954 & 0.157 & 7.21 & 20.9 & 2560; 3072 \\
L4 Triton & 53 & 88 & 58 & 349 & 0.0932 & 0.0813 & 0.0461 & 0.0870 & 5.47 & 20.9 & 2560; 3072 \\
A100 & 54 & 72 & 55 & 183 & 0.324 & 0.257 & 0.154 & 0.262 & 7.21 & 75.3 & at least 5120 \\
H100 & 33 & 43 & 34 & 103 & 0.417 & 0.326 & 0.194 & 0.327 & 6.96 & 75.2 & at least 5120 \\
\bottomrule
\end{tabular}
}A structure is one target with one seed and five samples. Times are medians over runs that compiled the model from scratch, and inference includes compilation. Cost is on demand unless marked Spot. The amortized cost spreads one compilation over the 8 targets of bucket 1024. Memory is the compiled peak plus the arguments. This metric counts the arguments twice on GPUs. Warm costs come from reruns on seed~1 of the 10 pilot targets, 2 of these targets at 1024. Supplementary Table~\Sref{S-tab:q2} has the full note.

\vspace{7\baselineskip}
\caption{\capTabSFuSixteen}\label{tab:app-hlo}
\centering\small\fitpage{\begin{tabular}{@{}lrrrrrrr@{}}
\toprule
Stage & Matrix products & With other input types & Conversions & Multiply & Select & Integer xor & Constants \\
\midrule
Input embedding & 34/34 & 5 & 37/34 & 71/73 & 8/8 & 164/164 & 185/159 \\
Templates & 31/31 & 0 & 132/132 & 145/147 & 72/76 & 123/104 & 307/287 \\
MSA module & 40/40 & 0 & 155/152 & 90/94 & 6/10 & 127/127 & 352/328 \\
\bottomrule
\end{tabular}
}We counted the operations in the program that XLA compiled for each chip (target 7U3J, seed 1). Each cell with two counts gives the v5e count, then the v6e count. ``With other input types'' counts the products whose operand or result types differ between the chips. The stages were assigned by a rule corrected after the first analysis was read. Supplementary Table~\Sref{S-tab:s-fu16} has the full note.

\end{table*}

\clearpage
\pdfbookmark[0]{Supplementary Material}{supplementbookmark}
\includepdf[pages=-,link,linkname=supplement]{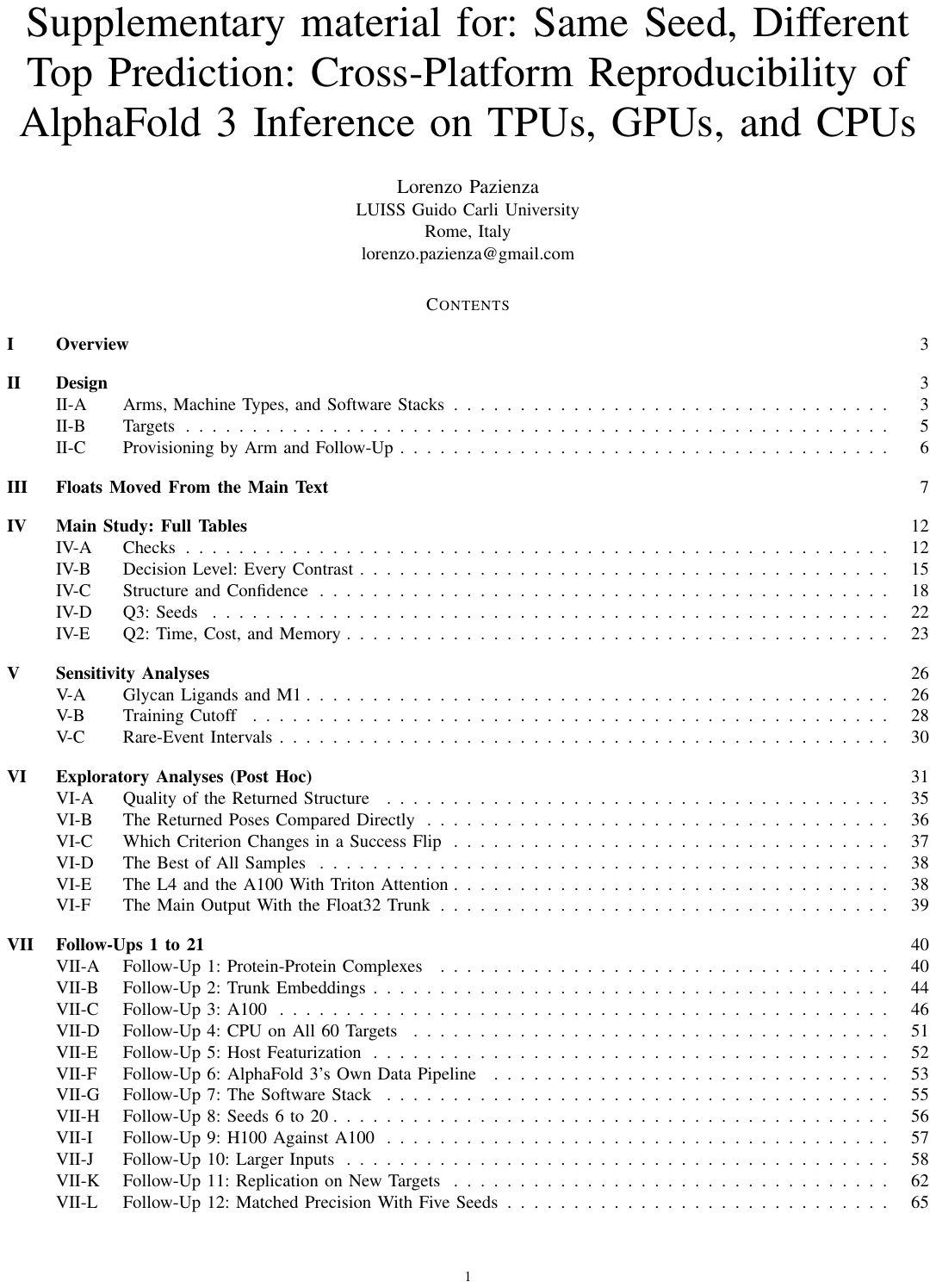}
\end{document}